# Friction coefficient and viscosity of polymer brushes with and without free polymers as slip agents

A. Gama Goicochea[1, 2,3*], R. López-Esparza[1, 4], M.A. Balderas Altamirano[1], E. Rivera-Paz[2], M. A. Waldo-Mendoza[2], E. Pérez[1]

[1]Instituto de Física, Universidad Autónoma de San Luis Potosí, San Luis Potosí, Mexico.

[2]Innovación y Desarrollo en Materiales Avanzados A. C., Grupo Polynnova, San Luis Potosí, Mexico.

[3]Tecnológico de Estudios Superiores de Ecatepec, Ecatepec de Morelos, Estado de México, Mexico

[4]Departamento de Física, Universidad de Sonora, Hermosillo, Sonora, Mexico

[*] Corresponding author. Electronic mail: agama@alumni.stanford.edu

# ABSTRACT

There is ample evidence that polymer brushes reduce friction between surfaces. Several industrial applications take advantage of this fact, such as those in plastic bag production, where the brushes act as slip agents; however, the complex mechanisms that give rise to such reduction of friction are not yet fully understood. In this work we report coarse grained, dissipative particle dynamics simulations carried out for surfaces functionalized with erukamide brushes, a polymer commonly used in the plastics industry as a slip agent between surfaces. We calculate their rheological properties, such as the coefficient of friction (COF) and the viscosity, η, as functions of the number of chains grafted on the surfaces under the influence of stationary, Couette flow. Moreover, we consider also the case when a fraction of the erukamide chains is not adsorbed and moves freely between the surfaces. We show that the COF reaches an equilibrium value of about 0.29 in these two cases, in agreement with experimental results. On other hand, the viscosity grows monotonically, as a result of the increasing collisions when the erukamide content is increased. The force between brushes is found to be in agreement with predictions from scaling theories. We find that the addition to free chains helps stabilize the film formed by the brushes and the solvent, as others have found experimentally. The mechanisms that give rise to these phenomena are studied in detail.

# I Introduction

Polymer brushes grafted on solid substrates are important for industrial applications due to their ability to reduce friction between the substrates [1 – 3], and because they can improve colloidal stability [4]. Additionally, biocompatible polymer brushes have been used to mimic the lubrication conditions found in bones and synovial joints [5 – 7]; in fact, these brushes can be used to obtain coefficients of friction as small as 0.001 [8]. Films made of polymers typically have large values of the coefficient of friction (COF), which leads to high energy consumption during their manufacture and affects their applications, such as in bag production. To reduce the COF different types of slip agents are added to the polymer matrix before the extrusion process takes place [9, 10]. One of those agents is erukamide [11], a fatty acid that has been used successfully for many years as a friction reducing and anti – block agent [12 – 14]. Ramirez and coworkers studied the reduction of the COF in surfaces covered with low density polyethylene as a function of the erukamide concentration [15, 16], finding that the COF is gradually reduced as the erukamide concentration is increased, reaching a plateau at about 0.5 µg/cm$^2$. This value is in agreement with that found by Molnar [17] in surfaces of a resin of polyethylene with amides. It is generally agreed that the amides are uniformly distributed in the polymer matrix immediately after extrusion and, as the polymer cools, they migrate to the surfaces, forming polymer brushes that reduce the COF when chemical equilibrium is reached. What is not usually recognized is that during this process some of the amides desorb and become segregated in between the surfaces, reaching an approximately constant concentration at chemical equilibrium. Several experimental investigations [18 – 20] have reported that polymer brushes couple with the free chains, leading to the stability of the film between the surfaces. However, the detailed mechanisms that give rise to the reduction of the friction coefficient are not fully understood yet, and numerical simulation can be a very useful tool in this respect. Among the techniques used to study the behavior of the COF in systems with polymer brushes are molecular dynamics [21, 22], Brownian dynamics [23], and more recently, dissipative particle dynamics (DPD) [24 – 26]. There are several studies of polymer brushes in contact with free polymer chains, using a variety of techniques [18 – 20, 27 – 31]. However, there are no reports to the best of our knowledge on the behavior of the COF in systems in which polymer chains are simultaneously forming brushes and segregate between surfaces, in the presence of the solvent. One exception is the work of Goujon and collaborators [26], but in their systems they reduced the number of grafted chains as the number of free chains was increased, resulting in a COF that grew with grafting density, in disagreement with experiments [15 – 17]. In this work, we use DPD to model two flat, parallel surfaces covered with erukamide chains forming polymer brushes, on the one hand, immersed in air as a solvent, and calculate the COF and the viscosity as functions of the density of polymers grafted on the surfaces, under the influence of stationary, Couette flow. On the other, we add a *fixed* concentration of erukamide molecules not forming brushes to the previously described system, to model the migration and segregation process of these molecules toward the surfaces, as chemical equilibrium is reached. This is a novel aspect of the work reported here. The motivation for this study is twofold. Firstly, there are numerous industrial applications where it is necessary that the friction between surfaces be as small as possible so that films can easily slide past each other. Additionally, much remains to be understood from the point of view of basic science concerning the mechanisms that give rise to low friction coefficients and/or low viscosity values in many body systems, starting from molecular interactions. With those aims in mind, in Section II we present the essentials of DPD, the details of the simulations and the systems we have studied. Section III is devoted to the presentation of our results and their discussion. Lastly, our conclusions are laid out in Section IV.

## II Models and Methods

The interaction model used here is DPD [32, 33], where the particles or beads represent sections of momentum carrying fluid with a given number of molecules, which defines the coarse graining degree. The reasons for choosing DPD over other computational techniques are the relatively large scales, both in size and time that can be reached with it, making the predictions more comparable with experiments on soft matter systems. Additionally, the pairwise nature of the DPD forces preserves the hydrodynamic modes of the fluid, which is a most important aspect for the work reported here. There are recent reviews that highlight the many successful applications of DPD to both equilibrium and non-equilibrium situations [34, 35], therefore we shall be brief here. The total force on the DPD beads is made up of the sum of three forces, namely a conservative force ($\boldsymbol{F}_{ij}^C$), a dissipative force ($\boldsymbol{F}_{ij}^D$), and a random one, ($\boldsymbol{F}_{ij}^R$), as expressed by equation (1):

$$\boldsymbol{F}_{ij} = \sum_{i \neq j}^{N} [\boldsymbol{F}_{ij}^C + \boldsymbol{F}_{ij}^D + \boldsymbol{F}_{ij}^R]. \tag{1}$$

The conservative force is given by a soft, linearly decaying repulsive function, shown in equation (2):

$$\boldsymbol{F}_{ij}^C = \begin{cases} a_{ij}(1 - r_{ij})\hat{\boldsymbol{r}}_{ij} & r_{ij} \leq r_c \\ 0 & r_{ij} > r_c \end{cases}, \tag{2}$$

where $\boldsymbol{r}_{ij} = \boldsymbol{r}_i - \boldsymbol{r}_j$, $r_{ij} = |\boldsymbol{r}_{ij}|$, $\hat{\boldsymbol{r}}_{ij} = \boldsymbol{r}_{ij}/r_{ij}$, $r_{ij}$ is the magnitude of the relative position between particles $i$ and $j$, and $a_{ij}$ is the intensity of the repulsion between those particles. This interaction constant depends on the coarse – graining degree [36]. The dissipative and the random forces are, respectively:

$$F_{ij}^D = -\gamma \omega^D(r_{ij})[\hat{r}_{ij} \cdot v_{ij}]\hat{r}_{ij} \quad (3)$$

$$F_{ij}^R = \sigma \omega^R(r_{ij})\xi_{ij}\hat{r}_{ij}, \quad (4)$$

where $\sigma$ is the noise amplitude and $\gamma$ is the friction coefficient and they are related as follows: $k_B T = \sigma^2/2\gamma$, where $k_B$ is Boltzmann's constant and $T$ the absolute temperature; $v_{ij} = v_i - v_j$ is the relative velocity between the particles, and $\xi_{ij} = \xi_{ji}$ is a random number uniformly distributed between 0 and 1 with Gaussian distribution and unit variance. The weight functions $\omega^D$ and $\omega^R$ depend on distance and vanish for $r > r_c$, and are chosen for computational convenience to be [33]:

$$\omega^D(r_{ij}) = [\omega^R(r_{ij})]^2 = max\left\{\left(1 - \frac{r_{ij}}{r_c}\right)^2, 0\right\}. \quad (5)$$

All forces between particles *i* and *j* vanish when their relative distance is larger than a finite cutoff radius $r_c$, which represents the inherent length scale of the DPD model and it is regularly chosen as the reduced unit of length, $r_c = 1$. The constants in equations (3) and (4) are chosen as $\sigma = 3$ and $\gamma = 4.5$, so that $k_B T = 1$, for all the cases reported in this work because those constants fix the thermostat, which should be the same for all types of brushes modeled here. In all our simulations we use a coarse – graining degree equal to three, which leads to $a_{ii} = 78$ for the interaction between particles of the same type; this sets the cutoff radius at $r_c = 6.46$ Å [36]. For the interaction between solvent particles and the monomers that make up the polymeric brushes, see Table A1 in the Appendix for details. The interaction parameter is reduced using $(k_B T/r_c)$. The systems modeled consist of monomeric solvent particles and linear chains (see Fig. 1(a)) made up of eight monomers joined by freely rotating harmonic springs:

$$F_{spring}(r_{ij}) = -\kappa_0(r_{ij} - r_0)\hat{r}_{ij}. \qquad (6)$$

In equation (6) the spring constant, $\kappa_0$, is set at $\kappa_0 = 100.0$ ($k_BT/r_c^2$) and $r_0 = 0.7r_c$ in all cases [37], except for the double bond in the erukamide chain, see Fig. 1(a), which is modeled with $\kappa_0 = 10.0$ ($k_BT/r_c^2$), see the Appendix for details. These chains are physically adsorbed on parallel surfaces by one of their ends (see Fig. 1(b)), and they can also be freely moving in between those surfaces. The simulation box is flanked by effective square surfaces placed at the ends of the box, perpendicularly to the $z$ – axis, i.e. at $z = 0$ and $z = L_z$. The force model for those surfaces is a simple, linearly decaying force law in the $z$ – direction, given by:

$$F_{wall}(z_i) = \begin{cases} a_w(1 - z_i/z_c)\hat{z} & z_i \leq z_c \\ 0 & z_i > z_c \end{cases}, \qquad (7)$$

where $a_w$ is the strength of the interaction between the DPD particles and the surfaces, chosen equal to $a_w = 70.0$ for the bead of the polymer chains adsorbed on the walls, and $a_w = 140.0$ for all other types particles; the cutoff distance is $z_c = 1.0$. We chose those values of $a_w$ to promote the adsorption of only the "heads" of the polymer chains on the surfaces, whose interaction with the walls is less repulsive (i.e., attractive) than the interaction of the rest of the beads in the chains with the walls. Only the conservative forces in equations (2), (6) and (7) are used in the calculation of thermodynamics properties, as it has been shown [38] that the dissipative and random forces (equations (3) and (4), respectively) do not contribute. All our simulations are performed in reduced units, with $(k_BT)^* = m^* = r_c^* = 1$, $\delta t^* = 0.01$. The time step $\delta t$ is reduced with $\delta t = (mr_c^2/k_BT)^{1/2}\delta t^*$, where $m$ is the mass of a DPD particle, while energy is reduced with $k_B T$. Using the mass of three water molecules per DPD particle, at room temperature, one obtains $\delta t \approx (6.3 \times 10^{-12} s)\delta t^*$. All simulations are performed in the canonical ensemble, at constant particle number ($N$), volume of the simulation box ($V$) and temperature, using periodic boundary conditions on the $xy$ – plane but not in the $z$ – direction. The positions and momenta of the particles are determined using the velocity Verlet algorithm adapted to the DPD model [39]. Non – equilibrium, stationary flow conditions were modeled by imposing a constant velocity ($v_0$) to the beads adsorbed on the surfaces, of equal magnitude but opposite direction with respect to the other surface, see Fig. 1(c).

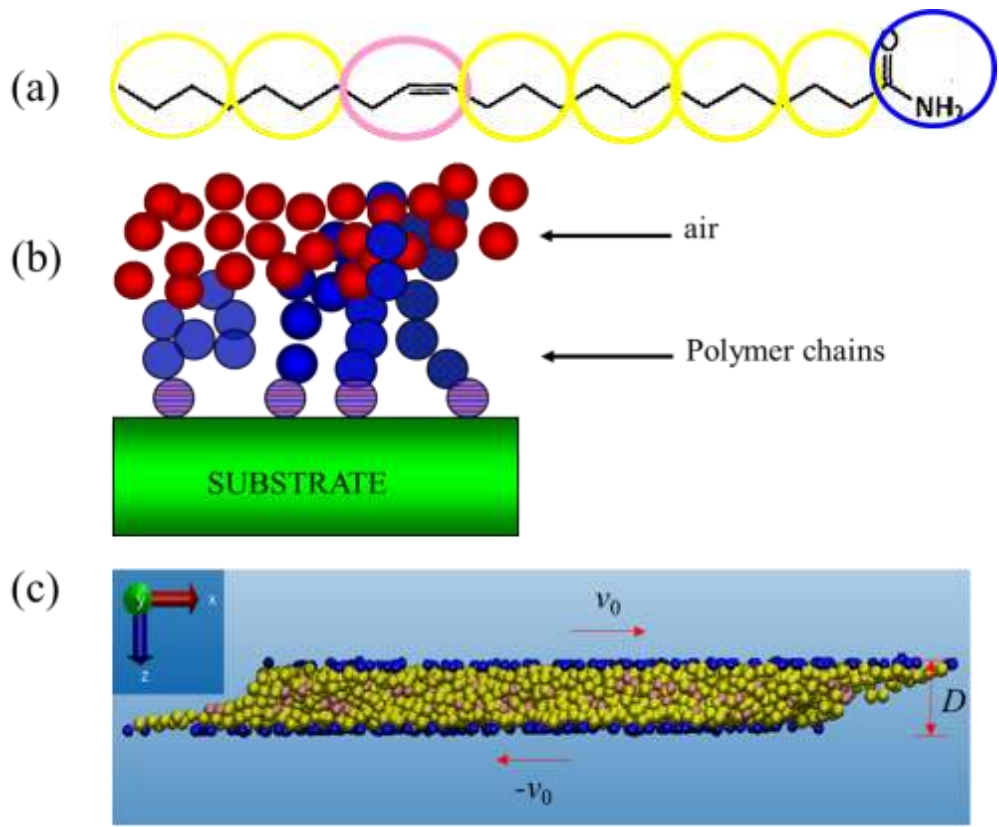

**Fig. 1**. (Color online) Schematic illustration of the setup used for our non – equilibrium simulations. (a) The coarse – grained model for erukamide used in this work. The circles represent the various types of DPD beads. The circle in blue represents the bead grafted to the surface. (b) Model for the erukamide brush; only one bead

per molecule is physically adsorbed on the surface. (c) A snapshot of the brushes under steady flow; $v_0$ is the shear velocity imposed on the adsorbed beads of erukamide; $D$ is the distance between the surfaces, which is kept constant. The solvent is omitted for clarity.

To predict values of the friction coefficient and viscosity one needs to perform non – equilibrium computer simulations where a steady external flow is applied to a confined fluid, as illustrated in Fig. 1(c). For the calculation of the COF ($\mu$) we used the equation $\mu = \langle F_x(\dot{\gamma})\rangle/\langle F_z(\dot{\gamma})\rangle$, see, for example, reference [40], where $F_x(\dot{\gamma})$ represents the magnitude of the force on the particles grafted onto each surface along the direction of the shear rate, $\dot{\gamma}$, and $F_z(\dot{\gamma})$ is the magnitude of the force on these particles, acting perpendicularly to the surfaces. The brackets indicate the time average of the forces. The viscosity ($\eta$) in the middle of the pore defined by the walls is obtained from the relation $\eta = \frac{\langle F_x(\dot{\gamma})\rangle/A}{\dot{\gamma}}$, where $A$ is the transversal area of the surface on which polymers are grafted [40]. The shear rate $\dot{\gamma}$ is equal to $2v_0/D$, where $D$ is the separation between the surfaces, which is kept constant. The factor of 2 in the shear rate arises from the fact that both surfaces are moving, rather than only one. Using this model of polymer brushes under the influence of an external flow (Couette flow) has been shown to lead to the correct prediction of scaling exponents, among other phenomena [41, 42]. Another advantage of DPD is that its thermostat, defined by the dissipative and random forces, remains stable even under non – equilibrium conditions [43]. The simulations were run for at least $10^2$ blocks of $2\times10^4$ time steps, with the first 40 blocks used to reach equilibrium and the rest used for the production phase. The dimensions of the simulation box in all cases were 44×44×4; the total density of the system (chains and solvent) was fixed at 3, and the shear velocity was set at $v_0$=1.0, all in reduced DPD units. Two sets of simulations were carried out; in one of them the number of polymer chains adsorbed per unit area (Γ) was increased, and the COF and viscosity were calculated for each value of Γ. In the other, a constant number of polymer chains (600) was added to the previous system, but these chains were not allowed to adsorb on the surfaces and form brushes.

## III Results and Discussion

We begin by presenting the results for the case where the erukamide chains are only forming polymer brushes. Figure 2(a) shows the COF and viscosity of the confined fluid as functions of the grafting density, where the value of the COF is seen to drop substantially at certain values of Γ, with η following qualitatively such non – monotonic behavior. In particular, raising the grafting density from Γ = 0.088 to Γ = 0.1875 leads to a drop in the COF from 0.30 to 0.02; further increasing of Γ to 0.20 increases the COF up to 0.38. When Γ is equal to or larger than 0.40 the COF behaves monotonically, reaching an equilibrium value of about 0.31.

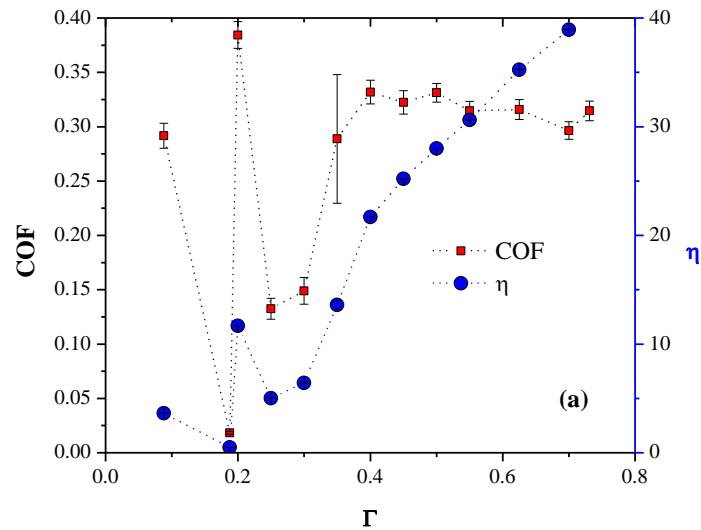

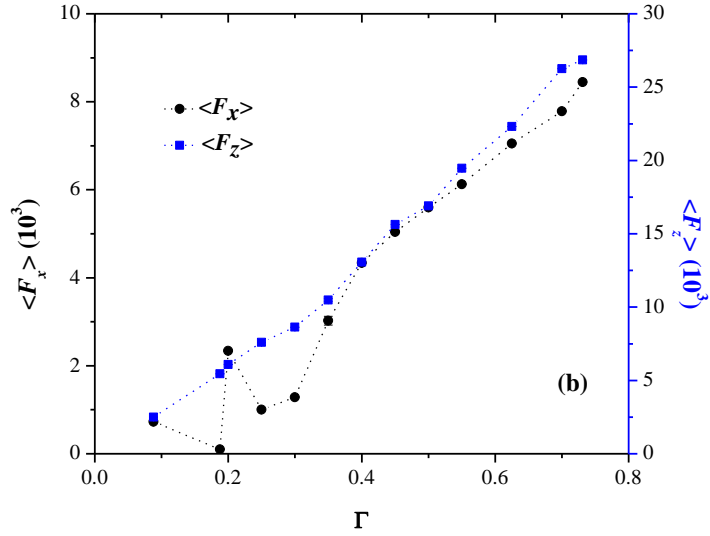

**Fig. 2.** (a) Coefficient of friction (COF, solid squares) and viscosity (η, solid circles) of two opposing linear polymer brushes as models of erukamide, as functions of the polymer grafting density (Γ). (b) The averaged forces along the flow direction ($\langle F_x \rangle$, solid circles) and perpendicularly to it ($\langle F_z \rangle$, solid squares), respectively, as functions of Γ. The dotted lines are only guides for the eye. All quantities are reported in reduced DPD units.

To trace the origin for such behavior in the rheological properties of the fluid we plot in Fig. 2(b) the average value of the forces along the direction of the flow ($\langle F_x \rangle$), and perpendicularly to it ($\langle F_z \rangle$). While the force along the $z$ – direction is seen to increase monotonically with Γ, the force along the flow direction shows fluctuations at low values of the grafting density. Since the COF is defined as the ratio of these forces, those fluctuations must be due to the fluctuation of $\langle F_x \rangle$ on Γ. It must be stressed that these are the resulting averaged forces measured on fully equilibrated systems over long periods; the simulations were monitored over 0.06 μs and the trends observed in Fig. 2 remained unchanged. For values of the grafting density equal or larger than 0.40 Fig. 2(b) shows that the mean force along the flow direction ($\langle F_x \rangle$) grows at approximately the same rate with Γ as does the mean force in the perpendicular direction, $(\langle F_z \rangle)$, which is why the COF remains approximately constant for $\Gamma \geq 0.40$, as shown in Fig. 2(a).

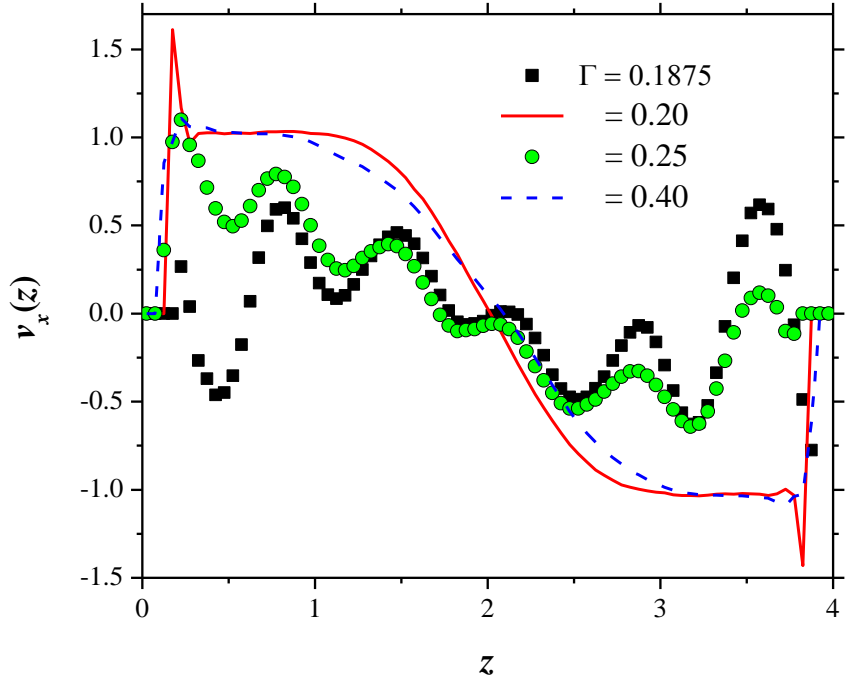

**Fig. 3.** (Color online) Profiles of the averaged component of the velocity along the direction of the flow of the erukamide beads not adsorbed on the surfaces (yellow beads in Fig. 1(a)), as a function of the direction perpendicular to the surfaces. The circles and squares represent cases where the COF is minimum while the lines correspond to grafting densities where the COF is maximum.

Additional quantitative information can be obtained from the velocity profiles, which are found in Fig. 3. Four values of $\Gamma$ are shown in Fig. 3, representing the average values of the $x$ – component of the velocity of the beads that are not adsorbed on the surfaces (yellow beads in Fig. 1(a)) along the direction perpendicular to the surfaces, $v_x(z)$. The solid and dashed lines represent the velocity profiles at grafting densities that produce relatively large values of the COF, see Fig. 2(a), while the symbols are cases where the COF is minimum. For the cases represented by lines one notices that in the middle of the pore the profiles are linear, as expected for Couette flow [40], and closer to the walls they reach the value of the shear velocity imposed on the surfaces, see Fig. 1(c). The velocity profiles shown by symbols in Fig. 3 correspond to grafting densities that lead to minima in the COF, and they clearly do not show a constant velocity gradient in the middle of the pore defined by the parallel walls. The oscillations seen in those profiles are due to the lack of uniformity in the formation of overlapping brushes on opposite surfaces, and they manifest themselves as minima in Fig. 2.

In Fig. 4(a) we present snapshots of DPD simulations of polymer chains under flow, see Fig. 1(a), forming brushes on parallel plane surfaces at four values of the chains' grafting density. Firstly, one distinguishes domains of erukamide adsorbed on the surfaces. The cases with $\Gamma = 0.1875$ and $\Gamma = 0.25$ correspond to the minima in the COF seen in Fig. 2(a). As Fig. 4(a) shows, for those two cases most of the chains forming brushes on one surface do not have an opposite brush on the same spot on the other surface. Therefore, there is no overlap between the brushes and the friction in those cases originates only from collisions between the chains with the solvent, and from chain – chain interactions within the brushes. That is the reason why the COF drops drastically at certain values of $\Gamma$ and it explains also the behavior of $\langle F_x \rangle$ in Fig. 2(b). On the other hand, when the brushes on one surface overlap with the brushes on the opposite surface, as seen in the snapshots for $\Gamma = 0.20$ and $\Gamma = 0.40$ in Fig. 4(a), the COF attains a larger and almost constant value, as observed in Fig. 2(a) for $\Gamma \geq 0.40$. In these cases, the viscosity grows monotonically with $\Gamma$, as a consequence of the entanglement between opposing brushes [24, 27]. Another salient feature of Fig. 4(a) is that a relatively small difference in $\Gamma$ can lead to markedly different values of the friction coefficient, e.g. when increasing $\Gamma$ from 0.1875 to 0.20 the COF goes from 0.01 up to 0.38, see Fig. 2(a). This is a consequence of physical adsorption of the chains on the walls, which allows the chains to move toward each other on the plane of the surfaces and adsorb as a cluster, thereby reducing their free energy.

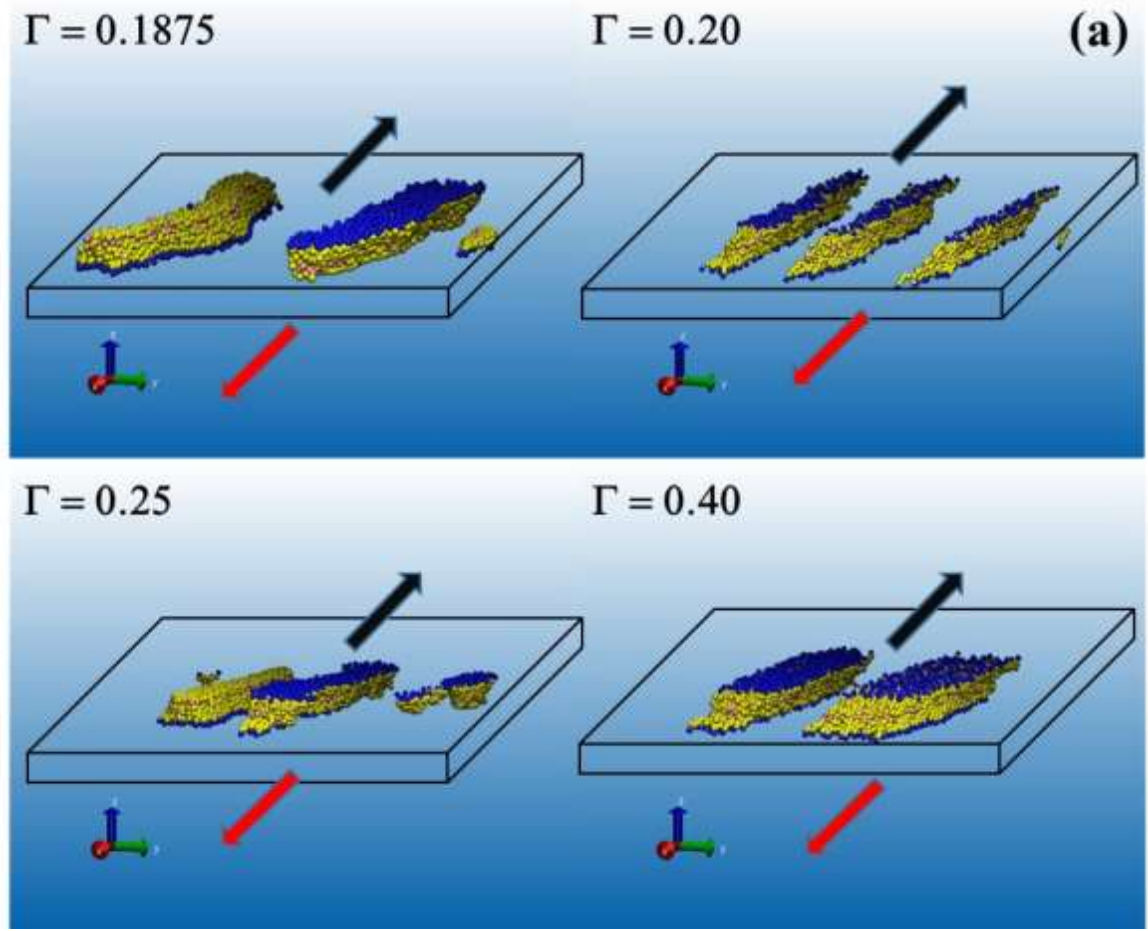

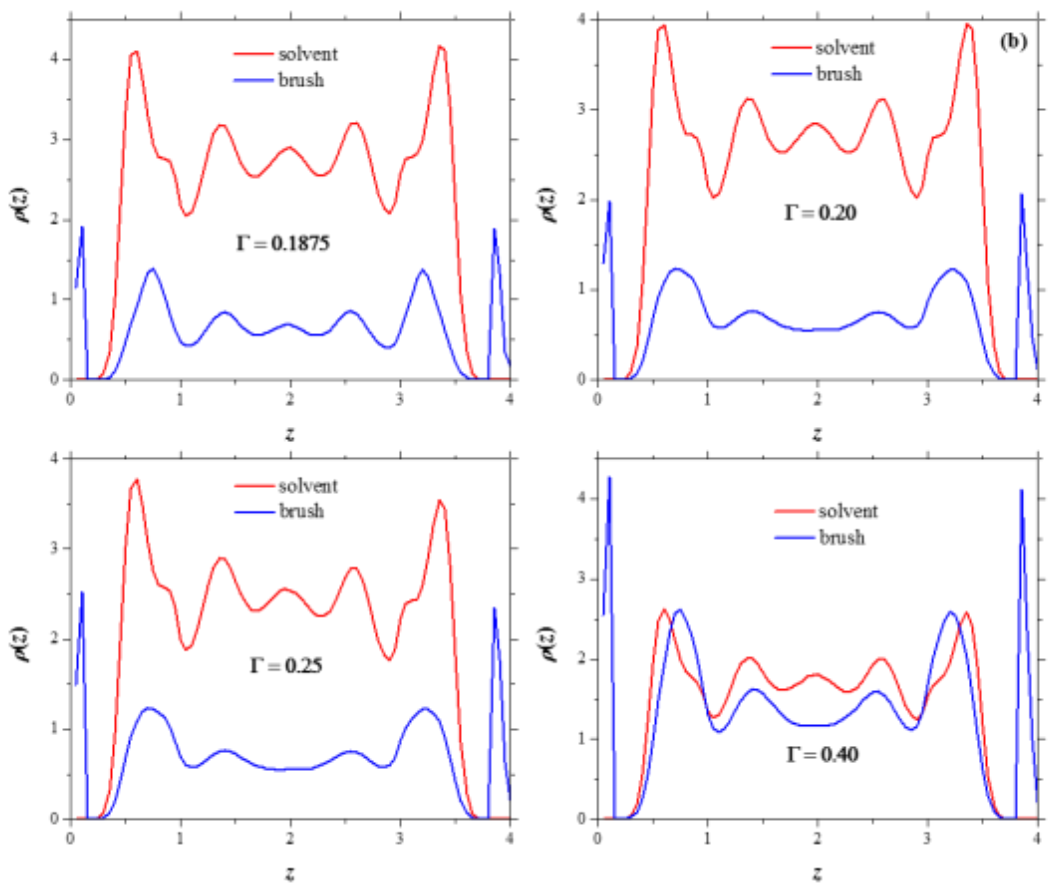

**Fig. 4.** (Color online) (a) Snapshots of the Couette flow simulations of erukamide brushes on parallel surfaces at four different values of erukamide chains per surface unit area, $\Gamma$. The blue beads represent the "heads" of the erukamide chains grafted to the surfaces (not shown); see Fig. 1(a) for the color code. The simulation box is illustrated by the parallelepiped in black line. The direction of the shear is indicated by the black (top face of the box) and red (bottom face of box) arrows. Notice that for the cases with $\Gamma = 0.1875$ and $\Gamma = 0.25$ most of the brushes on one surface do not have an opposing brush to overlap with on the opposite surface. The solvent was removed for clarity. (b) Density profiles of the brushes (blue line) and the solvent (red line) under flow at the four values of the grafting density shown in Fig. 4(a).

The density profiles corresponding to the four cases presented in Fig. 3 are shown in Fig. 4(b); the profiles of the solvent are shown in addition to those of the chains. The peaks closest to the surfaces represent the "heads" of the chains that are adsorbed onto them. It is important to note that the solvent penetrates the brush in all cases, although there are fewer solvent particles as the grafting density is increased. This is due to the relatively short distance between the surfaces; even at the largest grafting density shown in Fig. 4(b) the solvent's density profile has maxima close to the surface, indicating the penetration of solvent particles into the brush. As the erukamide grafting density is increased the solvent density must be reduced, to keep the total density constant, but even at relatively large values of $\Gamma$ the solvent penetrates the brush and in doing so it plays a lubricating role.

Let us now proceed to the case where there are free erukamide chains in addition to those forming brushes on the surfaces. Figure 5(a) shows the behavior of the COF and the viscosity in the middle of the pore as functions of the number of chains forming brushes per unit area. The number of free molecules was fixed in all cases at 600 erukamide chains. The COF shows markedly different behavior with respect to the one where there are only brushes, see Fig. 2(a); in particular, the COF starts at a relatively large value and then decreases until it reaches a plateau at about $\mu = 0.29$, which is comparable with that shown in Fig. 2(a). No oscillations are found when some of the erukamide molecules are allowed to desorb and move freely within the pore, in addition to the chains forming brushes. When $\Gamma$ is large the surfaces are covered uniformly by brushes, with the extra free chains overlapping with the brushes and the solvent, which leads to the same limiting value of the COF. The viscosity grows monotonically as $\Gamma$ is increased, reaching almost the same value as in the brush – only case at the comparable value of $\Gamma$. The maximum value of $\Gamma$ that can be reached is smaller than in the pure brush case because some of the chains are detached from the surfaces in this case and the global density is kept constant.

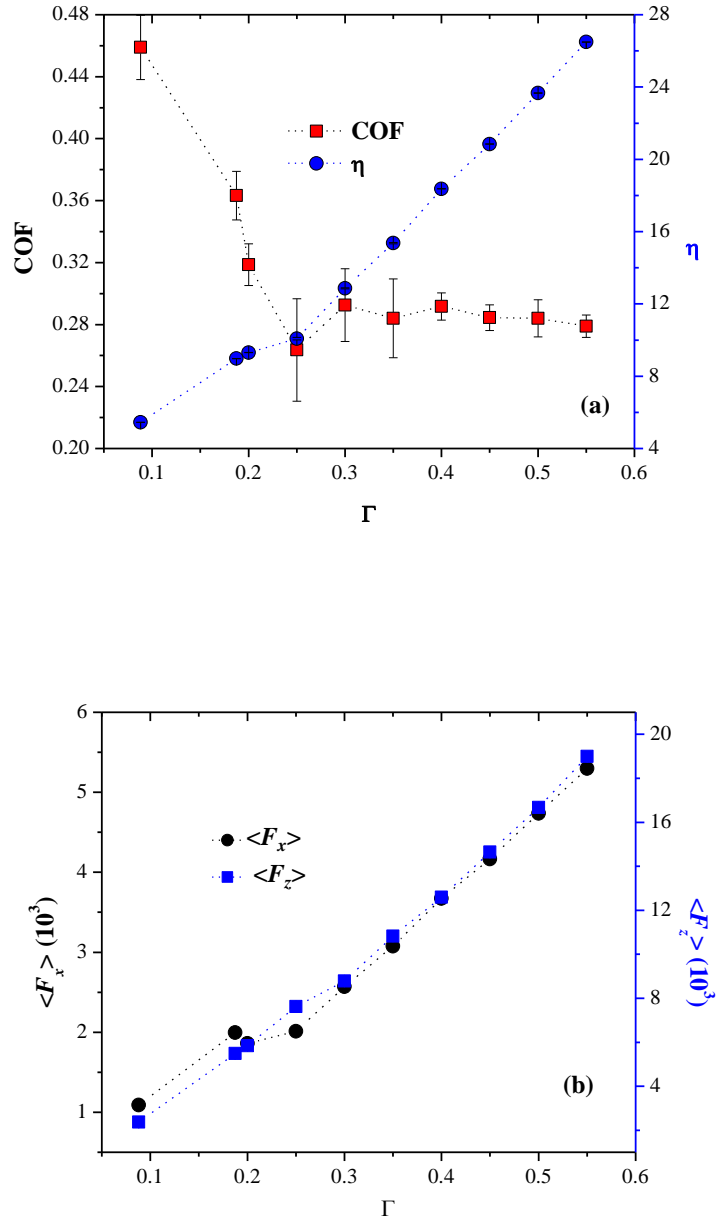

**Fig. 5.** (a) Coefficient of friction (COF) and viscosity ($\eta$) of two opposing linear polymer brushes as models of erukamide, as functions of the polymer grafting density ($\Gamma$). In addition to the brushes, there is a fixed concentration of free (not grafted) erukamide molecules, which is the same for all data shown here. (b) Average force along the flow direction (left $y$ – axis) and perpendicularly to it (right

*y* – axis). The dotted lines are only guides for the eye. All quantities are reported in reduced DPD units.

In Fig. 5(b) we present the average forces along the direction of the external flow, $\langle F_x \rangle$, and perpendicularly to it, $\langle F_z \rangle$, both showing monotonic increase in contrast with the trends seen in Fig. 2(b). At grafting densities larger than about 0.3 both forces increase with $\Gamma$ at approximately the same rate, which explains why the COF reaches a plateau. The increase in $\langle F_x \rangle$, see Fig. 5(b), is responsible for the equivalently similar behavior of the viscosity as $\Gamma$ is increased.

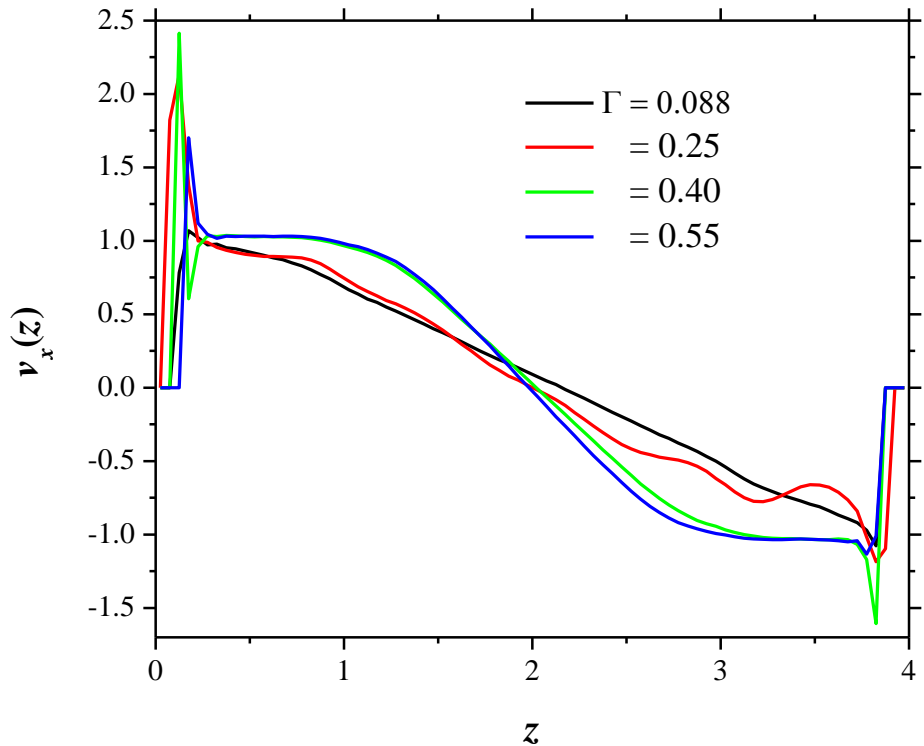

**Fig. 6.** (Color online) Velocity profiles along the direction of the flow of the beads that make up the erukamide chain (yellow beads in Fig 1(a)), corresponding to four different grafting densities. All chains are taken into account, i.e. those that are not adsorbed on the surfaces as well as those forming brushes.

The velocity profiles of some of the cases shown in Fig. 5 are presented in Fig. 6, in particular, the velocity profiles of the beads in yellow in Fig. 1(a), which are typical of Couette flow [34, 40, 42], as expected for polymer brushes under the influence of linear stationary flow. At small values of the grafting density, $\Gamma = 0.088$ and $\Gamma = 0.25$, one observes by extrapolating the velocity gradient in the center of the gap to the surfaces that the fluid composed of the beads that make up the brushes and the free chains also never reaches the velocity of the surfaces. This corresponds to the case of finite slip boundary condition [27, 44]. For larger grafting densities ($\Gamma = 0.40$ and $0.55$) the velocity profiles are very similar and show that the brush and free chain beads do reach the substrates' velocity at a distance of about $z \approx 1$ from each wall. These are examples of stick boundary conditions; the trends found here at small and large grafting densities are in agreement with those found with other methods [26, 27].

The snapshots in Fig. 7(a) show the formation of brushes entangled with those on the opposite surface, leading to the uniform behavior of the COF shown in Fig. 5(a). The extra chains, i.e., those that move freely between the surfaces become intertwined with the brush chains and they act as a homogenizing agent that contributes to the overlapping of the brushes from opposite surfaces. This is particularly clear for the snapshot corresponding to $\Gamma = 0.25$ in Fig. 7(a), which is to be compared with the snapshot at the same grafting density for brushes only, shown in Fig. 4(a). The addition of free chains promotes the association of the domains of adsorbed chains on opposite surfaces and that in turn leads to an increase in the COF, from about 0.01 for pure brushes (Fig. 2(a)), up to 0.36 when free chains are added, see Fig. 5(a). These trends are in agreement with those found by Yerushalmi – Rozen and Klein [18] who found that thin liquid films are stable when free chains are added to polymer brushes whereas without the free polymers the films dewet the surfaces. Even at relatively low grafting densities, e. g. $\Gamma = 0.25$ in Fig. 7, the free chains couple with the brush chains to enhance the stability of the films, as suggested by Safran and Klein [19].

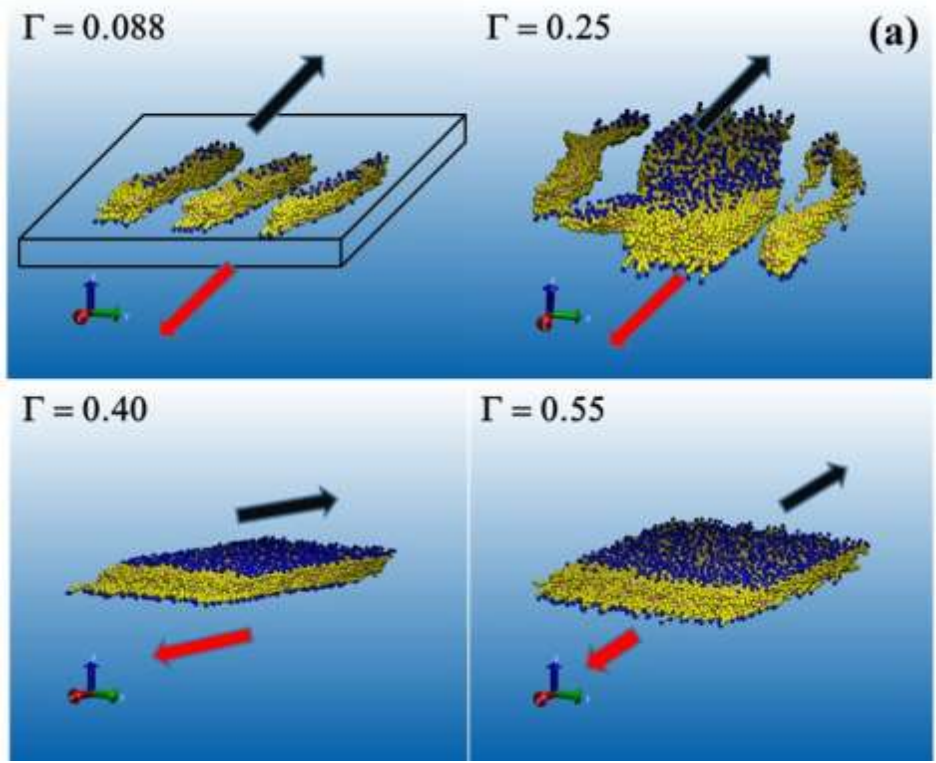

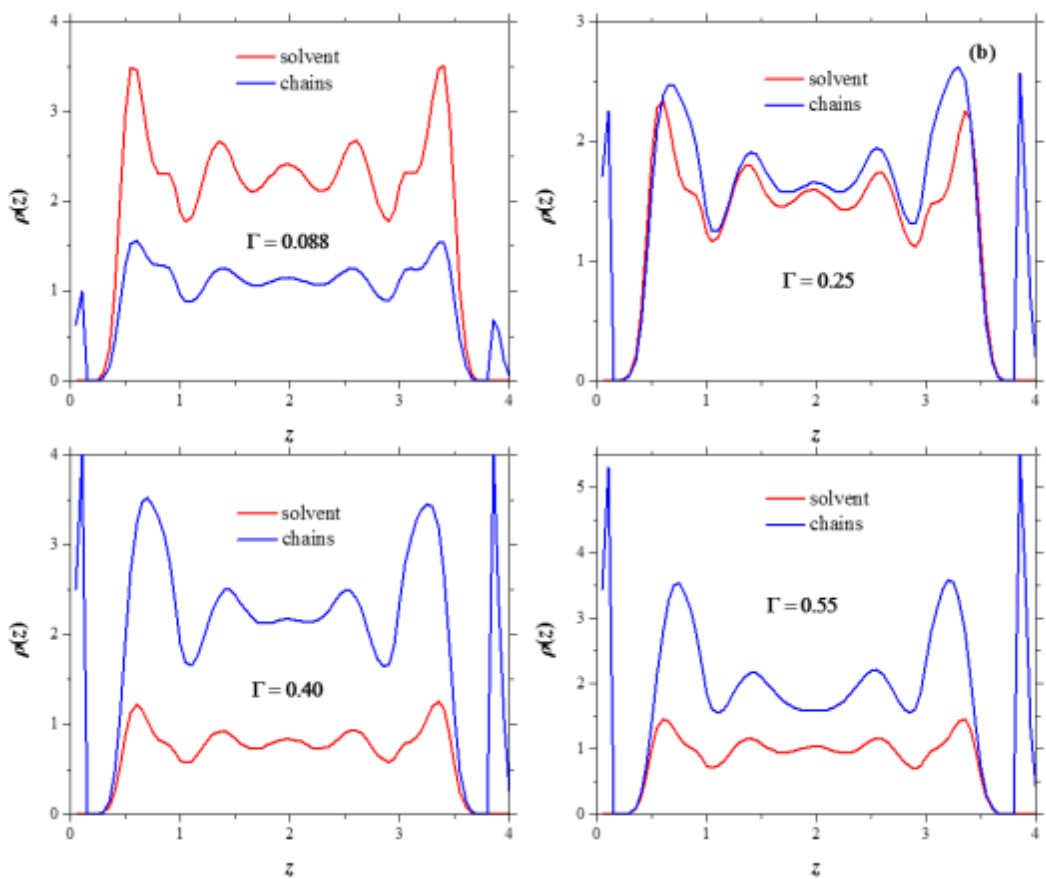

**Fig. 7.** (a) Snapshots of the simulations of erukamide chains moving freely within the surfaces in addition to those forming brushes, at four grafting densities. The simulation box is indicated by the parallelepiped in black line. The direction of the shear is indicated by the black (top face of the simulation box) and red (bottom face of the box) arrows. The free chains are entangled with the brushes. (b) Density profiles of two opposing linear polymer brushes as models of erukamide, as functions of the polymer grafting density ($\Gamma$). In addition to the brushes, there is a fixed concentration of free (not grafted) erukamide molecules, which is the same for all data shown here (600 chains in all cases). All quantities are reported in reduced DPD units.

The density profiles of the brushes plus free chains, see Fig. 7(b), are qualitatively similar to those of the system with brushes only, except the maxima in the chains profiles are wider and larger because they include the free chains as well. Comparing in particular the density profiles for $\Gamma = 0.25$ with those for the case with no free chains at the same grafting density, Fig. 4(b), shows that the free chains penetrate the brush, since the maxima are larger in Fig. 7(b). Moreover, the brush profile in the center of the pore ($z = 2$) does not show a peak of relative height larger than the rest of the maxima, as would be expected if the brush expelled the free chains to the center of the pore. It appears the free chains are homogeneously distributed within the brushes, since all the profiles in Fig. 7(b) are symmetrical, which keeps the COF at an almost constant value. A slight asymmetry is seen in the chains' density profile for $\Gamma = 0.25$, since the peak corresponding to the beads closest to the right wall (not the grafted beads) is somewhat larger than that corresponding to the beads closest to the left wall. This small asymmetry is also responsible for the slight asymmetry in the velocity profile corresponding to $\Gamma = 0.25$, see red line in Fig. 6. The solvent plays a lubricating role, as it penetrates the brushes even at the largest grafting densities, and this aspect determines to a large extent the value that the COF acquires, for it has been shown that implicit – solvent simulations of brushes and those of "dry" brushes predict larger values of the COF [45].

These trends compare reasonably well with experimental reports on erukamide brushes [15 – 17]. In particular, Ramirez and co – workers [16] found that polyethylene surfaces covered with erukamide showed a reduction of the COF as the erukamide concentration was increased, until the COF acquired an almost constant value (~ 0.27), in good agreement with our predictions, as shown in Fig. 5(a). In those experiments, the minimum COF is first obtained starting at an amide concentration of about 0.2 µg/cm$^2$, which translates into about 3 amide chains/nm$^2$, assuming all the amide chains at this concentration form brushes and do not desorb. In our simulations, this occurs when the erukamide concentration is 1.3 chains/nm$^2$, which suggests that in the experiments about half of the added amides form brushes and the rest desorb and associate in agglomerates, providing added lubrication to the surfaces. This coupling of the free chains with the brushes is found to promote the stability of the film, as seen by the behavior of the COF as a function of polymer grafting density ($\Gamma$) in Fig. 5(a), which reaches an approximately constant value at relatively small $\Gamma$, in agreement with various experimental studies [18 – 20].

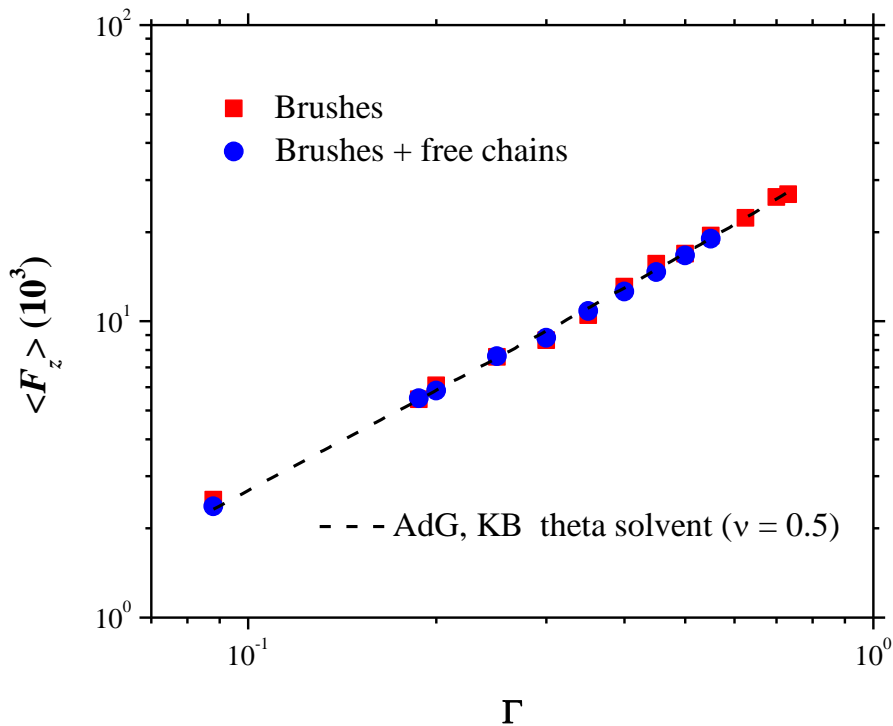

**Fig. 8.** Comparison of the average force perpendicular to the surfaces ($\langle F_z \rangle$) as a function of the brush grafting density ($\Gamma$) for the case when all polymer chains form brushes (solid squares), and when there are free chains in addition to brushes (solid circles). The dashed line is the fit to the Alexander – de Gennes (AdG) scaling equation of the osmotic pressure for brushes [46], which coincides with Kreer and Balko's (KB) prediction [47] under theta solvent conditions (our case here). All quantities are reported in reduced DPD units.

Lastly, in Fig. 8 we show the comparison of the average force perpendicular to the surfaces onto which the brushes are physisorbed for both cases studied here: when the erukamide chains all form brushes (solid squares in Fig. 8), and when a fraction of them is free and the rest is forming brushes (solid circles in Fig. 8). One notices that the forces almost overlap, in agreement with experiments on sheared brushes with and without free chains [20]. The dashed line in Fig. 8 represents the fit of the data to the Alexander – de Gennes (AdG) scaling of the osmotic pressure (*P*) between parallel plates separated by a distance *D*, covered with polymer brushes [46]:

$$P(D) = \frac{\langle F_z \rangle}{A} = (k_B T) f(a, D, N) \Gamma^y \tag{8}$$

where *A* is the area of the surfaces on which the chains are grafted; *f(a,D,N)* is a function that depends on the monomer size (*a*), polymerization degree (*N*), and *D* but does not depend on Γ. For fixed distance between the surfaces and polymerization degree (our case in this study), *f(a,D,N)* is a constant as a function of grafting density. The exponent *y* is related to the scaling exponent (ν) of the Flory radius $R_F$ of a polymer chain with *N* monomers, $R_F \sim N^\nu$ according to the relation [48]:

$$y = \frac{3\nu}{3\nu - 1}. \tag{9}$$

Under theta solvent conditions (our case), ν=0.5 and *y* = 3 [49]. The AdG scaling theory assumes a step – like density profile for the unperturbed brush and that the brushes do not interdigitate, which are known to be overly simplifying assumptions in actual polymer brushes [47, 50], see Figs. 4(b) and 7(b). In particular, in a recent report [47], Kreer and Balko develop a scaling theory for moderately compressed polymer brush bilayers, which allows for the interpenetration of the brushes. Their resulting force perpendicular to the surfaces can be written as follows [47]:

$$\frac{\langle F_z \rangle}{A} = (k_B T) g(a, D, N) \Gamma^{y'}, \tag{10}$$

where $g$ is a function that depends on the monomer size, distance between the surfaces and the polymerization degree, but when those three variables are constant, $g$ is also constant [47]. The exponent $y'$ of the grafting density in eq. (10) is defined as:

$$y' = \frac{2+5\nu}{3(3\nu-1)}. \tag{11}$$

Although the exponent defined by eq. (11) is different in general from that given by eq. (9), they both lead to the *same* result for brushes under theta solvent conditions ($\nu = 0.5$), namely $y = y' = 3$ [51]. Fig. 8 shows that the fit to the scaling is rather good; additionally, the force between brushes with or without free chains is almost the same. Although those scaling theories were develop for brushes in equilibrium, Deng et al. [44] have shown using also the DPD model that the density profile and equilibrium length of polymer brushes is not affected by the application of shear, regardless of grafting density. It can be concluded that the interpenetration of the brushes is not the leading factor affecting the force between brushes when the chain grafting density is increased, at constant intermediate compression and fixed shear rate. Other factors are expected to be more influential, such as the polymerization degree and the compression degree [47]. Although bond crossing between polymer chains is possible in the standard DPD model, the parameters used for the harmonic force that bonds neighboring beads in a chain, equation (6), have been tested to prevent such crossings. However, the erukamide chains we have modeled are too short to display reptation – like motion, therefore entanglement effects are not present here [25].

## IV Conclusions

The role played by erukamide chains as models for friction reducing agents has been studied here using mesoscopic scale computer simulations. It has been shown that the friction coefficient and viscosity η as functions of the number of chains grafted on the surfaces, under the influence of stationary Couette flow can be obtained reliably using this approach. The COF reaches an approximately constant value of about 0.31 for erukamide chains forming brushes, and 0.29 when free chains coexist with the brushes. These limiting values of the COF are found to be in agreement with experimental reports. The predicted increase in the average force perpendicular to the surfaces with grafting density, at fixed shear rate, can be accurately obtained from scaling theories of the osmotic pressure for polymer brushes, and it follows the same trend when free chains are added to the brushes. The role played by the explicitly included solvent and by the freely moving polymer chains is crucial not only for the lowering of the COF, but also for its reaching a stable value. Our work shows also that in the modeling of migrating polymer chains to the surfaces of a polymer matrix, as in the production of plastic bags, it is important to include desorbed, free chains simultaneously with those forming brushes. This provides a mechanism that promotes the stability of the film formed between the surfaces, which responds to the shear with an almost constant COF, through the coupling of the brushes and the free chains, in agreement with experimental reports [18 – 20]. Not adding free chains to the brushes may lead to the rupture of the film formed between them, which is detrimental for the intended role of erukamide as a slip agent; additionally the chains' desorption and liberation is an unavoidable and integral part of the chain migration process within the plastic matrix. Finally, we have shown that even if the film between the surfaces thickens with increasing grafting density (i.e., growing viscosity), that does not translate necessarily into a growing COF; this aspect has important applications in the polymer industry.

**Acknowledgements**

AGG acknowledges C. Ávila, J. D. Hernández Velázquez, and Z. Quiñones for discussions, and thanks the support of *A. Schulman de México*. MABA acknowledges PRODEP for a postdoctoral fellowship. AGG, MABA and RLE are indebted to IFUASLP for its hospitality, and thank J. Limón, for technical help.

**Appendix**

Here we provide full details of the conservative interactions chosen for the modeling of the erukamide brushes in air as the solvent. The nomenclature is defined in Fig.A1. The red beads (labelled as 1) represent monomeric solvent particles, while the erukamide chain is made up of eight beads joined by harmonic springs, see equation (6). Atomically detailed density functional theoretical calculations [52] have shown that the erukamide chains "bend" at the site of the double bond. The spring constants are $\kappa_0 = 100.0$ and $r_0 = 0.7$ in all cases, except for bead type 4, where $\kappa_0 = 10.0$ to incorporate the different nature of the bead that includes the double bond at the coarse grained level; its equilibrium position remains $r_0 = 0.7$. The interaction constant between the surface and the rest of the fluid, $a_w$ in equation (7), is $a_w = 70.0$ for the bead type 2; for bead types 1, 3 and 4 $a_w = 140.0$, in reduced DPD units. With these choices of interactions bead 2 is naturally adsorbed on the surfaces: although these interactions are repulsive, the one between the surface and bead 2 is less repulsive than those between the surface and the rest of the beads, which translates into an attractive interaction.

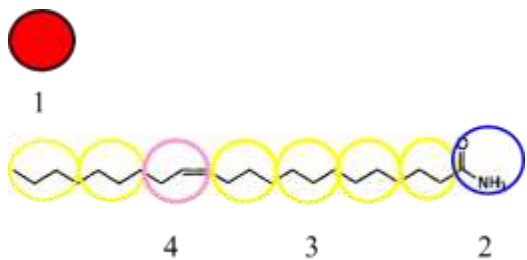

**Fig. A1.** (Color online) The labelling of the different types of DPD beads used in the simulations reported in this work. The (red) bead labelled 1 represents the solvent particles. Bead 2 (in blue) is the part of the erukamide chain that is adsorbed on the surfaces, while beads 3 and 4 constitute the brushes.

The values of the DPD conservative force constant between any pair of particles, $a_{ij}$ in equation (2) are presented in Table A1. The coarse graining degree used in this work and shown in Fig. A1 is such that each DPD bead has the equivalent volume of three water molecules, i.e., $V_b \approx 90\text{Å}^3$. To model appropriately the actual applications of erukamide as slip agents in plastic bag production, the solvent in this work is air. As is well known, the DPD equation of state is definite positive [33], therefore it does not allow liquid – vapor phase transitions. Although it is possible to modify the DPD equation of state so that it does predict liquid – vapor transitions [53], it is not necessary to do so here because in the problem that is the focus of this work there are no phase transitions of that kind: all simulations reported in this work were performed at constant density and temperature. There is only the coexistence of the polymer chains and the solvent particles, which can be adequately modeled as presented here.

**Table A1**. Matrix of the conservative interaction constant between DPD beads of different types, as defined in Fig. A1.

| $a_{ij}$ | 1 | 2 | 3 | 4 |
|---|---|---|---|---|
| 1 | 78.0 | 189.0 | 160.0 | 179.0 |
| 2 | | 78.0 | 78.0 | 85.0 |
| 3 | | | 78.0 | 85.0 |
| 4 | | | | 78.0 |

The values of the force constants, for interactions between particles of different type, presented in Table A1, were chosen keeping in mind that the erukamide chain has hydrophilic character when dissolved in water, hence when it interacts with air, the latter acts as a theta solvent for such polymer. For particles of the same type we followed the procedure laid out by Groot and Warren [36], using the Flory – Huggins model for the case when the DPD coarse – graining degree is three, which leads to $a_{ij} = 78$.

# GRAPHICAL ABSTRACT

# Friction coefficient and viscosity of polymer brushes with and without free polymers as slip agents

A. Gama Goicochea, R. López-Esparza, M.A. Balderas Altamirano, E. Rivera-Paz, M. A. Waldo-Mendoza, E. Pérez

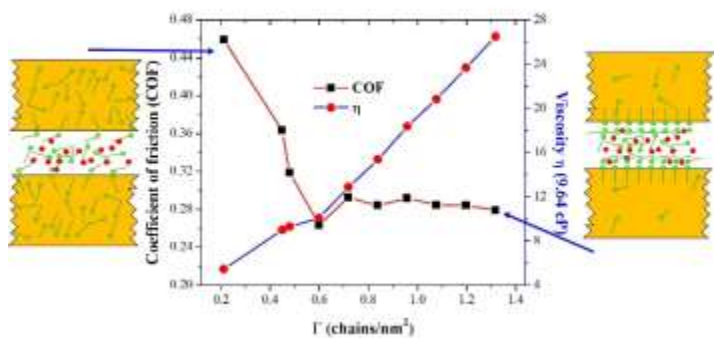